%% file: hpoXVI.tex
\documentstyle[aps,prl,twocolumn]{revtex}

\input{psfig}

\input{macroz.tex}

\begin{document}
\bibliographystyle{unsrt}
\newcommand{\be}{\begin{equation}}
\newcommand{\ee}{\end{equation}}
\newcommand{\bea}{\begin{eqnarray}}
\newcommand{\eea}{\end{eqnarray}}

\topmargin0cm         
\draft
\twocolumn[\hsize\textwidth\columnwidth\hsize\csname@twocolumnfalse%
\endcsname

\title
{Homoclinic Structure Controls Chaotic Tunnelling}

\author{Stephen C. Creagh$^{a,b}$\cite{nott} and Niall D. Whelan$^{a,c}$}
\address{
$^{a}$Division de Physique Th\'{e}orique\cite{cnrsbullshit},
Institut de Physique Nucl\'{e}aire 91406, Orsay CEDEX, France.\\
$^{b}$Service de Physique Th\'eorique,CEA/Saclay, 
91191, Gif-sur-Yvette CEDEX, France.\\
$^{c}$Department of Physics and Astronomy, McMaster University,
Hamilton, Ontario, Canada L8S~4M1.\\}

\date{\today}

\maketitle
\begin{abstract}
Tunnelling from a chaotic potential well is explained in terms of 
a set of complex periodic orbits which contain information about the 
real dynamics inside the well as well as the complex dynamics under the
confining barrier. These orbits are associated with trajectories which
are homoclinic to a real trajectory emerging from the optimal 
tunnelling path. The theory is verified by considering a model 
double-well problem.
\end{abstract}

]

Chaos is ubiquitous in nature. For this reason, its effect on quantum
systems is very well studied \cite{gutzwiller} and powerful
semiclassical techniques have been developed to treat it.  However,
tunnelling has largely remained impervious to such analysis.  We show
that a simple principle underlies the detailed behaviour of tunnelling
in chaotic systems --- state to state fluctuations in the tunnelling
rate are determined by a restricted set of complex classical
trajectories which can be constructed straightforwardly using known
techniques of dynamical systems.

Previous analyses of chaotic tunnelling exist in contexts where a
quantum state is initially localised in an integrable region of phase
space and tunnels into or through a chaotic region\cite{cat,SI}.  Our
goal is different; we consider systems that are almost completely
chaotic and tunnelling is through an energetic barrier. We then
construct a periodic-orbit theory of tunnelling analogous to
Gutzwiller's celebrated formula for the density of states. It relates
tunnelling rates to a set of complex classical trajectories which can
be understood intuitively as follows. In any potential barrier, there
is an optimal tunnelling route determined by a complex trajectory with
minimal imaginary action.  Interference effects in the tunnelling rate
are produced by trajectories which originate from and return to a
small region of phase space surrounding the minimal trajectory.  In
chaotic systems, such recurrent trajectories are understood on the
basis of ``homoclinic intersections''. Technical details come later,
but the main point is that such trajectories are standard in dynamical
systems theory and are relatively easy to find.  These orbits dominate
tunnelling rates in a manner similar to the way in which they control
wavepacket recurrences as discovered in \cite{SteveT}. See also
\cite{OdA}.

We treat double well potentials for which $V(-x,y)=V(x,y)$ and examine
the energy level splittings of quasi-degenerate doublets labelled by
$n$ (although our general conclusions apply equally to resonances in
metastable wells \cite{ourscarpaper,aletDelos}).  Detailed numerical
calculation will be for $V(x,y)=(x^2-1)^4+x^2 y^2+\mu y^2$ with
$\mu=1/10$. It will prove useful, instead of dealing directly with the
energy-level spectrum, to regard $q=1/\hbar$ as a parameter and ask
for the values $q_n$ for which a given energy $E$ is an eigenvalue ---
this leads to calculations with a fixed classical dynamics. We then
obtain doublets at values $q_n$ with splittings $\Delta q_n$ between
them. We investigate tunnelling by constructing the function
\cite{ourfirstPRL},
\begin{equation}\label{deff}
    f(q) = \sum \Delta q_n \delta(q-q_n).
\end{equation}
The spectrum and tunnelling splittings are encoded in $f(q)$.
Semiclassically, it can be approximated as a sum over {\it
pseudoperiodic} orbits which start on one side of the energy barrier
and, after evolution in complex time, finish at the symmetric image of
the initial point.  In \cite{ourfirstPRL} it was shown that when the
optimal tunnelling route is along a symmetry axis, a quasi-periodic
oscillation arises in the tunnelling rate which is related to the
action of a real periodic orbit, also on the symmetry axis. However,
there remain significant fluctuations which we now explain using
non-axial orbits homoclinic to this real orbit.

We can visualise a complex pseudoperiodic orbit as a one-dimensional
path in complex phase space. However, it is not unique --- it
corresponds to a given contour describing its complex time evolution;
any deformation of this contour gives an equivalent orbit. Because of
this freedom, a systematic and unambiguous description is difficult if
we work in full phase space. The situation simplifies considerably if
we restrict dynamics to a surface of section.  Pseudoperiodic orbits
then reduce to single, isolated points of intersection with the
surface of section and classification of them is simpler. An added
advantage is that metastable sytems can be treated within the same
formalism \cite{ourscarpaper,ournextpaper}.

Let $\Sigma$ denote a surface of section through one of the wells,
defined in the usual manner of real dynamics (eg. $x=x_0$, $\dot{x} >
0$ and $H=E$.) Let $\Trl:\Sigma\to\Sigma$ denote the first return
surface of section mapping, defined by letting trajectories start on
$\Sigma$ and evolve under real dynamics until they intersect $\Sigma$
once again. Real periodic orbits correspond to fixed points of some
iterate $\Trl^r$ of $\Trl$. We incorporate tunnelling by introducing a
second {\it complex} map $\Tim$ acting on the same section $\Sigma$ as
follows.

Any potential barrier has an orbit $\gamma_0$ which crosses it with
minimum imaginary action $iK_0$ in an imaginary time $-i\tau_0$. It
evolves in complex phase space except at two turning points where it
is fully real. There is a real point $z_0$ on $\Sigma$ which evolves
to one of these endpoints under real time evolution. It has a
symmetric image $\R z_0$ on $\R\Sigma$ which evolves to the other
endpoint. (Here $\R$ refers to the symmetry operation --- reflection
in $x$ in our example.)  The point $z_0$ evolves into $\R z_0$ if we
integrate along a complex time contour $C$ as follows: first the
contour evolves along the real axis until the trajectory reaches the
turning point of $\gamma_0$, then it descends parallel to the
imaginary axis until the turning point on the other side of the
barrier is reached and finally retraces the evolution parallel to the
real axis (in negative real time) until $\R z_0$ is reached. The net
time of evolution is $-i\tau_0$.

We define the map $\Tim$ by deforming $\gamma_0$. If a point $z'$ near
$z_0$ is chosen on $\Sigma$, a deformation of $C$ can be found such
that the final point $z''$ is on $\R\Sigma$. By invoking the symmetry
operation to map $\R\Sigma$ back to $\Sigma$, we define a complex
symplectic map $\Tim:z'\mapsto z=\R^{-1} z''$ from $\Sigma$ onto
itself. The central orbit $\gamma_0$ corresponds to a real fixed point
$z_0$ of $\Tim$ while other initial conditions, even if real, are
mapped to complex images under $\Tim$. Just as the real periodic
orbits give real fixed points of $\Trl^r$, the complex pseudoperiodic
orbits give complex fixed points of $\Tim\Trl^r$; such fixed points
define orbits with $r$ oscillations in one well before tunnelling
across the barrier.

Following \cite{bogomolnytrans}, the classical maps $\Trl$ and $\Tim$
have quantisations \cite{ourscarpaper} as unitary and hermitean
operators, respectively, acting on a Hilbert space which quantises
$\Sigma$.  The expression for $f(q)$ as a sum over the fixed points of
$\Tim\Trl^r$ has an interpretation as a sum over traces of such
operators, as will be shown in a future publication
\cite{ournextpaper}. Neglecting multiple tunnelling traversals and the
uniform analysis needed at the bottom and top of the wells,
\begin{equation}\label{rawsum}
	f(q) \approx f_0(q) + \frac{2}{\pi}\; \Re \sum_{r=1}^\infty 
        \sum_{\gamma \in P_r} 
			A_\gamma e^{iqS_\gamma} ,
\end{equation}
where $P_r$ denotes the set of fixed points of $\Tim \Trl^r$. The
$f_0(q)$ term is determined by $\gamma_0$, is monotonic and gives the
average behaviour of the splittings \cite{ourfirstPRL}. The
contributions with $r\geq 1$ have complex actions $S_\gamma$ with
nonzero real parts and describe fluctuations superimposed on $f_0(q)$.
The amplitudes are $A_\gamma = 1/\sqrt{-\det (M_\gamma-I)}$ where
$M_\gamma$ is the complex $2\times2$ symplectic matrix linearising
$\Tim\Trl^r$ about $\gamma$ (note that a different phase convention
was used in \cite{ourfirstPRL}.) The choice of root in $A_\gamma$ can
be unambiguously assigned using the real dynamics \cite{ournextpaper}.

In problems with an additional symmetry axis (for us it is along
$y=0$), the point $z_0$ is a fixed point of both $\Tim$ {\it and}
$\Trl$ and thus of each $\Tim \Trl^r$. The treatment in
\cite{ourfirstPRL} consisted of treating only this fixed point in each
of the terms in (\ref{rawsum}). We now systematically include the
other fixed points, which are generally complex.

\begin{figure}[h]
\vspace*{-4.5cm}\hspace*{-1.1cm} 
\psfig{figure=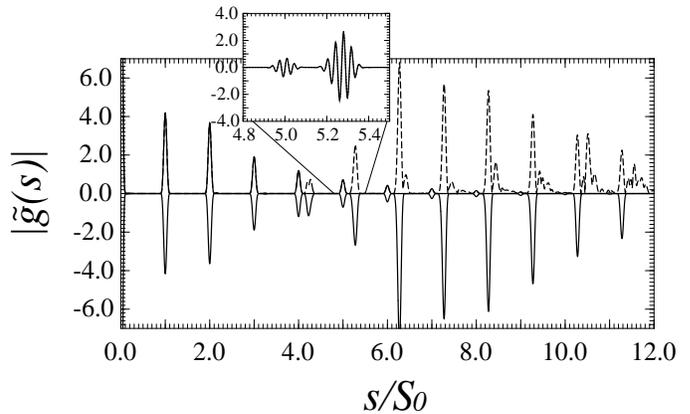,height=5.5in}
\vspace*{-4.0cm}
\caption
{\small Dashed curve: Fourier transform of $g(q)$ obtained quantum
mechanically. Upper solid curve: semiclassical prediction using just
the axial orbits. Lower solid curve: theory, using the six homoclinic
families shown in Fig. 2. Inset: the quantum-mechanical
$\Re\,\tilde{g}(s)$ and the semiclassical prediction in a limited
range (note there are two superimposed curves). Peaks not accounted
for by the lower curve correspond to uncomputed secondary
intersections.}
\label{FT}
\end{figure}

We now scale out the smooth exponential variation with $q$ by defining
$g(q) = f(q)/f_0(q)$ which is a function of order unity.  It is the
same as Eq.~(\ref{deff}) but with $\Delta q_n$ replaced by
$x_n=\Delta q_n/f_0(q_n)$. We find
\begin{equation}\label{defg}
  g(q) = 1 + \Re \sum_{r=1}^\infty 
        \sum_{\gamma \in P_r} 
			A_\gamma' e^{iq(S_\gamma-iK_0)} 
\end{equation}
where $A_\gamma'$ is the rescaled amplitude.  Note that $q$ only
appears in the exponents and that the residual action $S_\gamma-iK_0$
is predominantly real if $\Im \,S_\gamma \approx K_0$ so that $g(q)$
is an oscillatory function.

In Fig.~\ref{FT} we show the Fourier transform $\tilde{g}(s)$ of
$g(q)$ for energy $E=0.9$ and $q\in[30,100]$, chosen to put us well
within the semiclassical regime. For each $r$, the fixed point of
$\Tim\Trl^r$ at $z_0$ has action $S_\gamma=rS_0+iK_0$, where $S_0$ is
the action of the on axis real periodic orbit in one of the
wells. Therefore $\tilde{g}(s)$ exhibits a series of simple peaks at
the harmonics $s=rS_0$, well reproduced by the 
semiclassical theory.  More interesting is the additional structure,
primarily in the form of a sequence of regularly spaced peaks shifted
away from the simple harmonics, which we now explain using nonaxial
orbits.

Among barrier-crossing orbits, the minimum imaginary action is that of
$\gamma_0$; the contributions of other orbits are suppressed by order
$\exp(-\Delta K/\hbar)$ where $\Delta K$ is the additional imaginary
action. To be numerically significant, orbits in the sum
Eq.~(\ref{defg}) should have a small $\Delta K$ and so should begin
and end near $z_0$. On the other hand, to contain information about
the entire spectrum as implied in (\ref{deff}), they should fully
explore the phase space. These two requirements are satisfied by
trajectories homoclinic to $z_0$. Asymptotically, these trajectories
approach $z_0$ under both forward and backward time evolution as
$d(t)\sim e^{-\nu |t|}$, where $\nu$ is the stability exponent of the
central orbit and $t$ counts intersections with the Poincar\'e section.
In practice they are easily calculated as the intersections of the
stable and unstable manifolds of $z_0$ (the sets of points in
phase space which approach $z_0$ under forward and backward time
evolution, respectively.) For a general discussion see \cite{Wiggins}.

Let $(\cdots x_{-2},x_{-1},x_0,x_1,x_2\cdots)$ be the intersection of
one such trajectory with $\Sigma$.  Then a truncation $(x_{-M}\cdots
x_{-1},x_0,x_1,\cdots,x_N)$ is a finite length trajectory which is
almost a periodic orbit if $M$ and $N$ are large because $x_{-M}$ and
$x_N$ are exponentially close to $z_0$ and to each other. By slightly
perturbing the initial condition, one can find a nearby fixed point of
$\Tim \Trl^r$, where $r=M+N$, by complex Newton iteration.  The
tunnelling segment of this trajectory is very close to $\gamma_0$ and
as a result the imaginary part of the action is close to the optimal
value $K_0$. In the trace formula, there is then a competition of
exponentials --- the imaginary part of the action will decay
exponentially with length according to $\Im (S_\gamma-iK_0) \sim
e^{-\nu{\rm min}(M,N)}$ and this can be smaller than $\hbar$ so that
$e^{-q(S_\gamma-iK_0)}$ is nonnegligeable. For a fixed $r$, each
choice of $M$ (and $N=r-M$) gives a distinct fixed point although only
a subset will be numerically significant (the effective number depends
on $q$ and is given by the criterion that $\Im (S_\gamma-iK_0)
\sim e^{-\nu{\rm min}(M,N)} \lesssim \hbar$).  While the arguments
above are asymptotic in $M$ and $N$, in practice complex orbits could
usually be found as long as $M,N>2$ for the system we examine.

The near equal spacing of the peaks is because one extra iteration of
the real map amounts to including one extra $\Trl$-bounce near $z_0$;
to a good approximation this adds $S_0$ to the action. Also note that
the homoclinic peaks are larger than the peaks at $s=rS_0$ beyond
about $r=5$. Each homoclinic peak actually contains the contribution
of many fixed points, all with approximately the same action. This
quasidegeneracy is because there is more than one homoclinic
trajectory and because of the choice of $M$ for a given $r$ as
explained above. The amplitude of any single orbit initially grows
algebraically with length due to the increasing quasidegeneracy of the
orbit. This is ultimately overcome by the exponential decay with
length due to the instability of long orbits. In this case the maximum
is around $r=7$. There is additional structure which dominates at
larger $r$ and which we ascribe to secondary homoclinic
intersections. This scenario is like the calculation of wavefunction
recurrences in \cite{SteveT}, where there is also a selection of
trajectories which return to a classically small region of phase
space. There the region is determined by a Gaussian wavepacket,
whereas here it is by the near-Gaussian kernel of an operator
quantising $\Tim$ \cite{ourscarpaper,ournextpaper}. Fourier analysis
reveals peak structure similarly organised around homoclinic orbits
\cite{SteveT,heller}.

\begin{figure}[h]
\vspace*{-2.5cm}\hspace*{-1.0cm} 
\psfig{figure=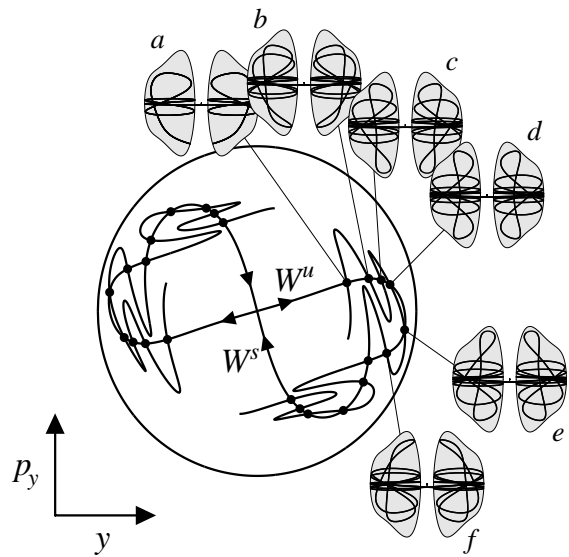,height=5.in}
\vspace*{-2.8cm}
\caption
{\small The surface of section defined by $E=0.9$ and $x=0.6$. The
circle is the boundary of the energetically allowed region. $W^u$ and
$W^s$ are the stable and unstable manifolds whose intersections define
the homoclinic trajectories: the six distinct ones being labelled $a$
to $f$. For each one, we show the trajectory in configuration space of
a corresponding tunnelling orbit.  Actually shown is the full periodic
orbit which is the double iteration of the pseudoperiodic orbit used
in calculation. Also, only the real parts are shown; the small
imaginary components being too small to see. }
\label{SOS}
\end{figure}

We now apply this formalism to the potential $V(x,y)$.  Shown in
Fig.~\ref{SOS} are the stable and unstable manifolds of $z_0$ in the
$x=0.6$ surface of section. Of particular interest is a sequence of
$6$ primary intersections labelled $a$ to $f$.  (All other
intersections shown are either iterates of these six or are related by
reflection in $y$ and $p_y$.) This is a rather rich structure; often
there are only $2$ primary intersections \cite{Wiggins}. Each of these
intersections defines a homoclinic trajectory, and a corresponding
series of fixed points of $\Tim\Trl^r$. We show the projections in
real configuration space of complex trajectories defined for each of
these intersections by the truncation $(M,N)=(3,5)$ corresponding to
the peak at $s\approx 8.2S_0$.

The symmetries of time-reversal and reflection in $y$ lead to
degeneracy among the orbits. The families $b$ and $c$ are respectively
mapped into those of $f$ and $e$ under a combination of the two
symmetries. Congruent but distinct families are obtained if we apply
either symmetry alone. The family $a$ maps to itself under
time-reversal but a distinct family is obtained under reflection in
$y$, while the converse applies to $d$. We therefore only have four
inequivalent families: $a$, $b$, $c$ and $d$ with degeneracies 2, 4, 4
and 2 respectively.  This high degeneracy (including also the choice
$M$), compared to the single axial orbit, is responsible for the
relative dominance of the associated peaks in Fig.~\ref{FT}.  In the
resolution presently available, the individual contributions of
$\{a,b,c,d\}$ are not resolved, but presumably would be with a wider
$q$-window.

To get detailed agreement in Fig.~\ref{FT} we had to include a family
of ``ghost'' orbits in addition to the orbits described above.  Near
the intersections $a$ and $b$, there is a switchback in the stable
manifold emanating from $f$ that almost produces two additional
intersections; at lower energies these intersections actually occur
and lead to two new families of orbits. These can be tracked up to
$E=0.9$, although they become complex. (These orbits are analogous to
the ghost orbits of \cite{Kusetal}.) One of them is exponentially
large and is excluded through the Stokes phenomenon. The other does
contribute and is an important component in the leading edges of the
peaks. The Fourier transform is then almost completely reproduced up
to about $s=8S_0$.  Thereafter additional side peaks emerge which can
be explained by the secondary intersections (which would be seen in
Fig.~\ref{SOS} if the invariant manifolds were extended.) Therefore, a
systematic calculation of homoclinic structure suffices for a complete
understanding of chaotic tunnelling.

While the details presented here are limited to systems with
additional symmetry in $y$, we claim that the guiding principle
applies generally. If the real extension of $\gamma_0$ is not
periodic, the tunnelling is given by trajectories homoclinic
to it, though they are not as simply organised as here. The
important physical notion is that homoclinic trajectories are
asymptotic to the optimal tunnelling route while fully exploring
classical phase space.

There has been significant recent effort to understand dynamical
tunnelling in which the barriers are not energetic but come from
dynamical effects such as KAM surfaces \cite{cat,SI}. In particular,
an exhaustive semiclassical analysis was carried out for one
particular system in \cite{SI}. The authors considered the propagation
of a quantum state initially localised in a regular region of phase
space and used complex trajectories to describe its penetration into a
chaotic region. Explicit semiclassical formulas were developed but it
is difficult to make a direct comparison to our results. Here we have
considered a purely spectral quantity in terms of classical invariants
--- the complex periodic orbits --- while they considered the
propagation of a specific quantum state. In terms of the quantities
considered, namely splittings, the work of \cite{cat} is closer in
spirit. However, because of the complicated, mixed nature of the phase
space, there are no explicit semiclassical results of the type derived
here. In particular, the splittings typically vary over many orders of
magnitude, even after normalisation by the local mean, unlike what we
observe. While it would be desireable to develop an explicit
semiclassical formalism for such systems, it is far from obvious how
this could be done.

The problem of the statistics of the normalised splittings will be
addressed in a later publication \cite{ournextpaper}. For now we
remark that they are not governed by the Porter-Thomas distribution
but depend on the specific properties of $\gamma_0$. This is in spite
of the fact that splittings are formally similar to resonance widths
for which one expects Porter-Thomas \cite{porter}. This effect is
quite general and should also apply to certain resonance problems.

\end{document}

%% file: macroz.tex
\def\opket#1#2 {  {#1} |{#2}\rangle }

\def\det{{\rm det}\,}

\def\Im{{\rm Im}}

\def\K2{{\cal K}}

\def\R{{\sigma}}
\def\Re{{\rm Re}}

\def\Tim{{\cal F}}

\def\Trl{F}
